\documentclass[twocolumn,epsfig,graphics,noshowpacs,floatfix,nofootinbib]{revtex4-1}

\usepackage{amsmath,amsfonts,amssymb,graphics,graphicx,epsfig,color,times,bbm}
\usepackage{amsthm}
\usepackage{psfrag}
\usepackage{braket}
\AtBeginDocument{\usepackage{booktabs}}
\graphicspath{{figures/}}
\usepackage{array}
\newcolumntype{P}[1]{>{\centering\arraybackslash}p{#1}}
\newcolumntype{M}[1]{>{\centering\arraybackslash}m{#1}}
\usepackage{graphicx}
\usepackage{mathtools}
\usepackage{soul}
\usepackage[normalem]{ulem}
\usepackage{url}
\usepackage{placeins}

\begin{document}

\title{Comparison of schemes for highly loss tolerant photonic fusion based quantum computing}

\author{Quantum Architecture @ PsiQuantum}

\date\today

\begin{abstract} 

We summarize the performance of recently-proposed methods for achieving fault tolerant fusions-based quantum computation with high tolerance to qubit loss, specifically aimed at photonic implementations. 
\end{abstract}

\maketitle

Fusion based quantum computing (FBQC) is an approach to universal quantum computing in which projective entangling measurements (`fusions') are made between fixed-size entangled `resource states' which are arranged to form a \emph{fusion network}.
FBQC was introduced in \cite{FBQC} and its application to fault tolerant quantum computing was further developed in \cite{DBA,FusionComplexes, bell2023optimizing, paesani2023high, sahay2023tailoring, pankovich} which explore a variety of fusion networks and different approaches to adaptive fusion measurements. FBQC is particularly well suited for photonic quantum computing using dual-rail encoded photonic resource states and linear optical fusion measurements \cite{BrowneRudolph}. The error tolerance of fusion networks is intrinsically linked to the size of the resource states. In this paper we summarize the landscape of published schemes for fault tolerant FBQC with dual-rail photonic qubits under a common metrics of footprint cost and loss threshold, we analyze additional fusion networks and then compare them against the fundamental limits of loss tolerance.

Loss is the primary error affecting photonic qubits. The \emph{loss per photon threshold} (LPPT) is the maximum tolerable loss in an error model where every photon of each resource state experiences optical loss with the same probability. To increase tolerance to both photon loss and fusion failure \cite{FBQC} introduced the idea of performing the fusions transversally across encoded qubits in the fusion network and reported that for a \{2,2\}-encoded 6-ring resource state\footnote{ The notation \{n,m\} is used to denote a Shor code with n X-repetitions and m Z-repetitions. This is also sometimes known as a quantum parity check code.} the LPPT was $2.7\%$ when the fusion failure basis was randomized. This is a factor of 3.4 improvement over the LPPT achieved using boosted fusion and no encoding (0.79\%). This is not unexpected - boosting adds more photons for modest improvements in success probabilities, and has no intrinsic loss tolerance.

\begin{figure}[t]
    \centering
    \includegraphics[width=0.91\columnwidth]{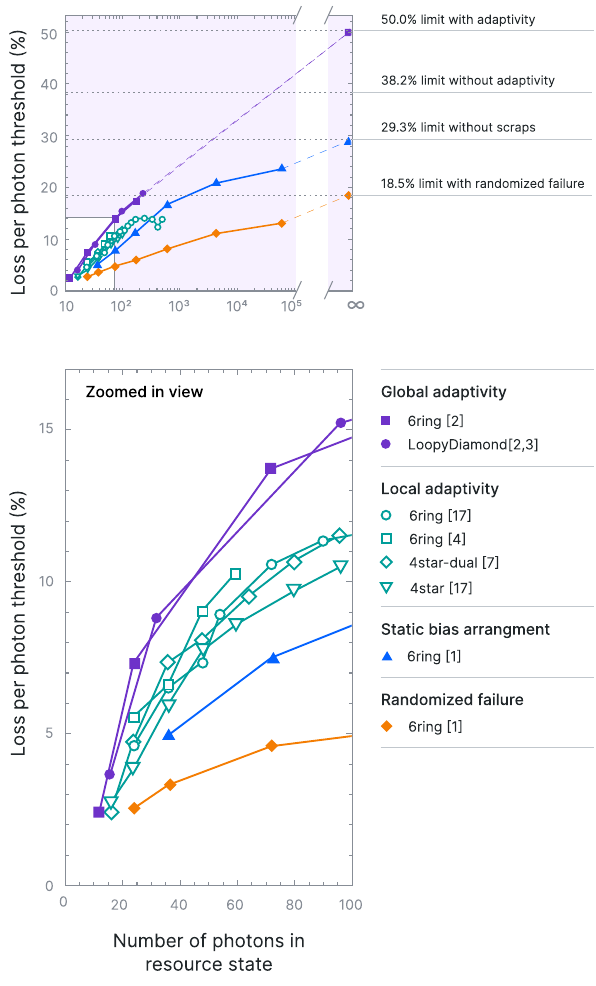}
    \caption{\textbf{Loss tolerance of published FBQC schemes} Loss per photon threshold versus number of photons in various encodings of 4-star, 4-star dual, 6-ring and 8-loopy-diamond fusion networks. The data used is given in Table~\ref{table:data}. \textbf{Top}: Limiting loss tolerance behavior for different types of measurements. Adaptive schemes can (under the right configuration) achieve 50\% loss threshold in the infinite resource size limit. \textbf{Bottom}: Zoom in on the region of interest with resource state sizes of fewer than 100 qubits. Color series distinguish different approaches to fusion configuration. In some cases data points were omitted to show only the bounding envelope of best performing points for each scheme. Solid points show PsiQuantum's results.}
    \label{fig:plot}
\end{figure}

The LPPT can be further increased by using \emph{adaptivity} to exploit the biased nature of a linear optical fusion measurement. This involves staggering the physical fusions and allowing feedforward to modify the fusion failure bases depending on the outcomes of the previous fusions.  Methods for increasing loss tolerance using adaptivity with encoded states and measurements have been studied \cite{DBA, varnava2006loss, azuma2015all, li2015resource, lee2015nearly, ewert2016ultrafastprl, lee2019fundamental, hilaire2021error, bell2023optimizing, lee2023parity, pankovich} in a wide variety of settings. Specific to FBQC, Bell et al.~\cite{bell2023optimizing} showed that adaptivity can be used to achieve a LPPT of 5.7\% with a 24 qubit resource state which is a 6-ring with a four-qubit encoding.

In \cite{DBA} a form of adaptivity in the FBQC setting termed \emph{dynamic bias arrangement} was introduced. This adaptivity can be divided into two parts - `local adaptivity' in which feedforward is used within a locally encoded fusion, and `global 
\FloatBarrier

\noindent adaptivity' in which feedforward is used at the fusion network level (i.e. at the scale of the topological code). Making use of only local adaptivity increases the LPPT to 5.7\% (without boosting). But the two types can be combined for maximum effect. The particularly effective adaptive method in \cite{DBA} termed \emph{exposure based adaptivity}  optimizes the choice of fusion bases for physical fusions (which may include single qubit measurements in the high loss regime) in a topological fusion network such that information which is more important for suppressing the growth of large erasure clusters is prioritized. For the \{2,2\}-encoded 6-ring resource state this increases the LPPT to  $7.5\%$. These methods can be straightforwardly applied to larger local encodings leading to increased performance. For example using the $\{7,4\}$-encoded 6-ring fusion network they yield a LPPT of $17.4\%$. 

More recently Pankovich et al.~\cite{pankovich} and Song et al.~\cite{songetal} proposed other locally adaptive methods. ~\cite{songetal} found that for the \{2,2\}-encoded 6-ring resource state their method reaches a LPPT of $4.8\%$, and for the $\{7,4\}$-encoded 6-ring fusion network a LPPT of $13.97\%$. ~\cite{pankovich} uses a `4-star dual' fusion network with encoded Bell pair resource states and found a LPPT of $5.0\%$ for a 24-qubit resource state.

Higher thresholds can be achieved with larger resource states, and it is natural to wonder what the limits of the LPPT are in various settings. In the non-adaptive case,~\cite{lee2015nearly, ewert2016ultrafastprl, lee2019fundamental, hilaire2023linear} have shown that the limit of loss tolerance is $29.3\%$ when a single photon lost in a fusion results in all information being erased. However, in~\cite{losslimitspaper} we show even without adaptivity an LPPT of $38.2\%$ ($\approx (3-\sqrt{5})/2$) can be achieved by accounting for the non-stabilizer information (known as scraps) which may still available in the presence of loss. In the adaptive case with single photon measurements, the limit of the LPPT is $50\%$. 

These limits of loss tolerance are all achievable but require extremely large (though still fixed size) resource states. For practical quantum computing, it is therefore important to account for the complexity of making the initial resource state when trying to compare schemes.

The simplest such \emph{footprint metric} is the total number of photons in the resource state. Figure~\ref{fig:plot} contains a summary of of optical FBQC schemes analyzed to date which make use of adaptive fusions on encoded qubits. There is a strong relationship between the number of photons in a resource state and the loss per photon threshold. It can also be seen that the introduction of increasingly powerful adaptivity improves the tradeoff such that higher threshold can be achieved with smaller resource states over the entire range.

\begin{table}
\centering
\vspace{20pt}
\begin{tabular}{cccc} 
\hline
\textbf{Resource state} & \textbf{Size} & \textbf{Local encoding} & \textbf{Cost}\\ 
\textbf{} & \textbf{(\#qubits)} & \textbf{} & \textbf{(\# 3GHZs)}\\ 
\hline\hline
4-star &16 & \{2,2\} & 256  \\ 
\hline
6-ring &24 & \{2,2\} & 1520\\ 
\hline
8-LD & 32 & \{2,2\} & 1120 \\ 
\hline
4-star &112  & \{7,4\} & 12928\\ 
\hline
6-ring &168 & \{7,4\} & 66560 \\ 
\hline
8-LD  &224 & \{7,4\} & 52480 \\
\hline
\end{tabular}
\caption{Costing of resource states using  3GHZ  initial states and 2-way (unboosted) fusions, making the optimistic assumptions of efficient multiplexing and lossless fusion as outlined in the Appendix. Different preparation schemes could significantly change the resource costing, but we expect the conclusion that the relative cost of a resource state goes up very significantly with the size of local encoding to be robust. }\label{table:resources}
\end{table}

Counting the number of photons in a resource state is a limited footprint metric. If the resource states are made from fusions on smaller entangled states (e.g. 3-GHZ states) then it turns out that resource states with the same number of qubits may require differing numbers (on average) of these smaller states to prepare. Moreover, the numbers depend strongly on assumptions of the preparation methods available. To illustrate this we outline a simplified such calculation in the Appendix, and summarize the results in Table~\ref{table:resources}. This includes the example of a 32-qubit resource state which is cheaper to prepare (and with a higher threshold) than the 24-qubit \{2,2\}-encoded 6-ring resource state. 

Also complicating the analysis of good FBQC schemes is that with the \emph{fusion complex} construction of \cite{FusionComplexes} the number of possible fusion networks exploded - over 600 instances of fusion networks were identified. Furthermore, since a Shor-code is a surface code, any Shor-encoded fusion complex is automatically a fusion complex with a modified geometry of resource states and fusions. Moreover; schemes to make use of $n$-way fusions (boosted or unboosted) for $n>2$ were proposed as were non-surface-code constructions giving an even larger scope of fault tolerant fusion networks. 

Some of these new fusion networks are competitively performing even if we restrict to standard 2-way transversal fusions between Shor-encoded qubits. For example, by using a $\{2,2\}$-encoded version of the `cuboctahedral' fusion complex of \cite{FusionComplexes}, along with the adaptivity methods of \cite{DBA}, we find an improved threshold of $9.0\%$ (vs. $7.5\%$ for the comparable 6-ring fusion network). We refer to this as the \emph{loopy diamond} fusion network. This scheme uses a resource state with 8 qubits (8-LD) and therefore is not directly comparable to the 6-ring fusion network. However, our analysis shows that the 8-LD state in fact can have a lower preparation overhead (see Table~\ref{table:resources}).

These schemes are still not optimal. Syndrome information and other state quality metrics can also be fruitfully used at multiple intermediate scales, all the way up to the global scale in which all prior measurement information is utilized to decide subsequent fusion structures~\cite{bombin2024fault}. 
However the analysis of this plethora of possibilities ultimately only makes sense in conjunction with more realistic resource accounting and error modeling.

\textbf{Authors:} Sara Bartolucci, Tom Bell, Hector Bomb\'in, Patrick Birchall, Jacob Bulmer, Christopher Dawson, Terry Farrelly, Samuel Gartenstein, Mercedes Gimeno-Segovia, Daniel Litinski, Yehua Liu, Robert Knegjens, Naomi Nickerson, Andrea Olivo, Mihir Pant, Ashlesha Patil, Sam Roberts, Terry Rudolph, Chris Sparrow, David Tuckett, Andrzej Veitia.


\bibliographystyle{apsrev4-1} 
\bibliography{fbqcNEW}

\appendix\label{app:resources}
\section{Resource state costing}

As a simple footprint metric we count the ``number of 3GHZ states" that would be required to make the target resource state using destructive Type-II fusions under the following assumptions: 
\begin{itemize}
    \item We treat each fusion as succeeding with probability $1/2$, ignoring the probability reduction coming from loss in the photon lifetime.
    \item Each step with a probability $p$ requires an overhead of $1/p$. This ignores any inefficiency introduced by the switching networks~\cite{bartolucci2021switch}.
    \item Any state which is the result of fusion failure is thrown away. 
\end{itemize}

Under these assumptions the cost of a state which is a result of $n$ successful type-II fusions between two states of cost $C_1$ and $C_2$ is $(C_1 + C_2)2^n$. Resource state generation proceeds in a series of steps creating states of increasing size. Any state which can be represented as a tree-like ZX diagram (e.g. the resource states for the loopy diamond fusion network) can have $n=1$ for all steps but other states with loops (e.g. 6-ring resource states) will have $n>1$ in some steps.

The resource state cost is heavily dependent on the order in which fusions are performed and the best representation of the resource state. We use a heuristic optimization routine to minimize the cost to find the numbers in Table~\ref{table:resources}. This heuristic is close to optimal for tree-like resource states. A similar optimization was presented in ~\cite{lee2023graph}.

There are some caveats to this approach. The resource state costs are dependent on the optimization heuristic, which can be slightly sub-optimal for the 6-ring network (exact optimization is inefficient, but close-to-optimal strategies can be performed with reasonable computational resources). To gain further confidence on the quality of the heuristic, we can show that the resource state cost for a state of size $S$ is lower bounded by $(S-2)^2$ for the setting described above. This is an achievable lower bound for some tree-like states and is close for tree-like states in general. As an example, for the \{7,4\} encoded 8-LD resources state the heuristic cost is $52480$ whereas the lower bound is only 7\% lower at $49284$.
We also note that in general it is possible to make resource states more efficiently if resource states that result from failure are recycled and not thrown away~\cite{kieling2007minimal, gross2006potential}.

Despite these caveats and assumptions, Table~\ref{table:resources} does emphasize that the relative cost of a resource state goes up very significantly with the size of local encoding, and we expect this to remain true irrespective of the preparation method details.

The ``number of photons per encoded fusion'' as advocated in \cite{songetal} is not a good metric for resource costing. In fact two identical schemes can appear to have different costs, while two very different schemes can appear to have identical costs! This follows because: 
\begin{itemize}
    \item[i)] It is arbitrary which level of the concatenation is counted as `the encoded qubit', for example a $\{2,2\}$-encoded 4-star state is the same as a $\{2,1\}$-encoded 8-qubit state. These two interpretations yield identically performing schemes but would count by such a metric as having a different cost. 
    \item[ii)] The number of photons per encoded fusion does not take any account of the size of the resource state. A $\{2,2\}$-encoded 4-star or a $\{2,2\}$-encoded 100-star would be counted as having the same cost.\footnote{ A minor point: in Figure~3 of \cite{songetal} the $x$-axis is conflating two quantities. For the points labelled EFBQC, the $x$-axis corresponds to the success probability of encoded fusions while the points labelled as FBQC correspond to the physical fusion success probability. The encoded fusion success probability will go up with the size of the physical fusion but the physical fusion success probability will not. The definition of FBQC always incorporated local encodings and has never made boosted fusion a requirement.}
\end{itemize}
   
The ``Number of 3GHZ states'' metric overcomes these issues by removing any dependence on the interpretation of the encoding level, and accounting for preparation cost in a simplified model. However this metric still has the limitations of not accounting for loss in the success probability, being restricted to specific primitive operations and requiring optimization of the preparation routine which is not guaranteed to be optimal. Furthermore the resource state cost must be weighed against the spread of Pauli errors in the preparation routine, which can only be fully analyzed in the context of the fusion network. Improved resource costing metrics would be a valuable tool for comparing FBQC protocols without the need to account for physical errors in their full complexity.

\renewcommand{\arraystretch}{0.8}

\begin{table*}[]
\begin{tabular}{@{}llllllll@{}}
\toprule
\textbf{Adaptivity   methods} & \textbf{Ref.}                & \textbf{Fusion network} & \textbf{\begin{tabular}[c]{@{}l@{}}Unencoded \\ Resource \\ state\end{tabular}} & \textbf{\begin{tabular}[c]{@{}l@{}}Local \\ encoding\end{tabular}} & \textbf{\begin{tabular}[c]{@{}l@{}}Number of\\ qubits in \\ resource state\end{tabular}} & \textbf{\begin{tabular}[c]{@{}l@{}}Loss per \\ photon   \\ threshold\end{tabular}} & \textbf{\begin{tabular}[c]{@{}l@{}}Boosted \\ fusion?\end{tabular}} \\ \midrule
Exposure based adaptivity     & \cite{DBA,FusionComplexes}   & LoopyDiamond            & 8-LD                                                                            & \{2, 1\}                                                           & 16                                                                                         & 3.9\%                                                                              & unboosted                                                           \\
Exposure based adaptivity     & \cite{DBA,FusionComplexes}   & LoopyDiamond            & 8-LD                                                                            & \{2, 2\}                                                           & 32                                                                                         & 9.0\%                                                                              & unboosted                                                           \\
Exposure based adaptivity     & \cite{DBA,FusionComplexes}   & LoopyDiamond            & 8-LD                                                                            & \{4, 3\}                                                           & 96                                                                                         & 15.4\%                                                                             & unboosted                                                           \\
Exposure based adaptivity     & \cite{DBA,FusionComplexes}   & LoopyDiamond            & 8-LD                                                                            & \{7, 4\}                                                           & 224                                                                                        & 18.8\%                                                                             & unboosted                                                           \\
Exposure based adaptivity     &\cite{DBA}   & 6ring    & 6ring    & \{2, 1\}  & 12                                                                                         & 2.6\%                                                                              & unboosted                                                           \\
Exposure based adaptivity     & \cite{DBA}   & 6ring                   & 6ring                                                                           & \{2, 2\}                                                           & 24                                                                                         & 7.5\%                                                                              & unboosted                                                           \\
Exposure based adaptivity     & \cite{DBA}   & 6ring                   & 6ring                                                                           & \{4, 3\}                                                           & 72                                                                                         & 13.9\%                                                                             & unboosted                                                           \\
Exposure based adaptivity     & \cite{DBA}   & 6ring                   & 6ring                                                                           & \{7, 4\}                                                           & 168                                                                                        & 17.4\%                                                                             & unboosted                                                           \\
Local adaptivity              & \cite{pankovich} & 4star dual              & \{2,1\}-BP                                                                      & \{2,2\}                                                            & 16                                                                                         & 2.6\%                                                                              & unboosted                                                           \\
Local adaptivity              & \cite{pankovich} & 4star dual              & \{2,1\}-BP                                                                      & \{2,3\}                                                            & 24                                                                                         & 5.0\%                                                                              & unboosted                                                           \\
Local adaptivity              & \cite{pankovich} & 4star dual              & \{2,1\}-BP                                                                      & \{2,4\}                                                            & 32                                                                                         & 5.7\%                                                                              & unboosted                                                           \\
Local adaptivity              & \cite{pankovich} & 4star dual              & \{2,1\}-BP                                                                      & \{3,3\}                                                            & 36                                                                                         & 7.5\%                                                                              & unboosted                                                           \\
Local adaptivity              & \cite{pankovich} & 4star dual              & \{2,1\}-BP                                                                      & \{4,3\}                                                            & 48                                                                                         & 8.3\%                                                                              & unboosted                                                           \\
Local adaptivity              & \cite{pankovich} & 4star dual              & \{2,1\}-BP                                                                      & \{4,4\}                                                            & 64                                                                                         & 9.7\%                                                                              & unboosted                                                           \\
Local adaptivity              & \cite{pankovich} & 4star dual              & \{2,1\}-BP                                                                      & \{5,4\}                                                            & 80                                                                                         & 10.9\%                                                                             & unboosted                                                           \\
Local adaptivity              & \cite{pankovich} & 4star dual              & \{2,1\}-BP                                                                      & \{6,4\}                                                            & 96                                                                                         & 11.7\%                                                                             & unboosted                                                           \\
Local adaptivity              & \cite{pankovich} & 4star dual              & \{2,1\}-BP                                                                      & \{7,4\}                                                            & 112                                                                                        & 12.2\%                                                                             & unboosted                                                           \\
Local adaptivity              & \cite{bell2023optimizing}      & 6 ring                  & 6 ring                                                                          & 4-qubit OGC                                                        & 24                                                                                         & 5.7\%                                                                              & boosted                                                             \\
Local adaptivity              & \cite{bell2023optimizing}      & 6 ring                  & 6 ring                                                                          & 6-qubit OGC                                                        & 36                                                                                         & 6.8\%                                                                              & unboosted                                                           \\
Local adaptivity              & \cite{bell2023optimizing}       & 6 ring                  & 6 ring                                                                          & 8-qubit OGC                                                        & 48                                                                                         & 9.2\%                                                                              & unboosted                                                           \\
Local adaptivity              & \cite{bell2023optimizing}       & 6 ring                  & 6 ring                                                                          & 10-qubit OGC                                                       & 60                                                                                         & 10.5\%                                                                             & unboosted                                                           \\
Local adaptivity              & \cite{songetal}  & 4star                   & 4 star                                                                          & \{2, 2\}                                                           & 16                                                                                         & 2.9\%                                                                              & unboosted                                                           \\
Local adaptivity              &  \cite{songetal}  & 4star                   & 4 star                                                                          & \{2, 3\}                                                           & 24                                                                                         & 4.0\%                                                                              & unboosted                                                           \\
Local adaptivity              &  \cite{songetal}  & 4star                   & 4 star                                                                          & \{2, 4\}                                                           & 32                                                                                         & 4.4\%                                                                              & unboosted                                                           \\
Local adaptivity              &  \cite{songetal}  & 4star                   & 4 star                                                                          & \{3, 3\}                                                           & 36                                                                                         & 6.1\%                                                                              & unboosted                                                           \\
Local adaptivity              &  \cite{songetal}  & 4star                   & 4 star                                                                          & \{4, 3\}                                                           & 48                                                                                         & 7.9\%                                                                              & unboosted                                                           \\
Local adaptivity              &  \cite{songetal}  & 4star                   & 4 star                                                                          & \{5, 3\}                                                           & 60                                                                                         & 8.8\%                                                                              & unboosted                                                           \\
Local adaptivity              &  \cite{songetal}  & 4star                   & 4 star                                                                          & \{6, 3\}                                                           & 72                                                                                         & 9.1\%                                                                              & unboosted                                                           \\
Local adaptivity              &  \cite{songetal}  & 4star                   & 4 star                                                                          & \{5, 4\}                                                           & 80                                                                                         & 9.9\%                                                                              & unboosted                                                           \\
Local adaptivity              &  \cite{songetal}  & 4star                   & 4 star                                                                          & \{6, 4\}                                                           & 96                                                                                         & 10.7\%                                                                             & unboosted                                                           \\
Local adaptivity              &  \cite{songetal}  & 4star                   & 4 star                                                                          & \{7, 4\}                                                           & 112                                                                                        & 11.4\%                                                                             & unboosted                                                           \\
Local adaptivity              &  \cite{songetal}  & 4star                   & 4 star                                                                          & \{10, 4\}                                                          & 160                                                                                        & 12.8\%                                                                             & unboosted                                                           \\
Local adaptivity              &  \cite{songetal}  & 4star                   & 4 star                                                                          & \{9, 6\}                                                           & 216                                                                                        & 11.9\%                                                                             & unboosted                                                           \\
Local adaptivity              &  \cite{songetal}  & 4star                   & 4 star                                                                          & \{17, 4\}                                                          & 272                                                                                        & 14.0\%                                                                             & unboosted                                                           \\
Local adaptivity              &  \cite{songetal}  & 4star                   & 4 star                                                                          & \{14, 6\}                                                          & 336                                                                                        & 13.3\%                                                                             & unboosted                                                           \\
Local adaptivity              &  \cite{songetal}  & 6ring                   & 6 ring                                                                          & \{2, 2\}                                                           & 24                                                                                         & 4.8\%                                                                              & unboosted                                                           \\
Local adaptivity              &  \cite{songetal}  & 6ring                   & 6 ring                                                                          & \{2, 3\}                                                           & 36                                                                                         & 6.7\%                                                                              & unboosted                                                           \\
Local adaptivity              &  \cite{songetal}  & 6ring                   & 6 ring                                                                          & \{4, 2\}                                                           & 48                                                                                         & 7.5\%                                                                              & unboosted                                                           \\
Local adaptivity              &  \cite{songetal}  & 6ring                   & 6 ring                                                                          & \{3, 3\}                                                           & 54                                                                                         & 9.1\%                                                                              & unboosted                                                           \\
Local adaptivity              &  \cite{songetal}  & 6ring                   & 6 ring                                                                          & \{4, 3\}                                                           & 72                                                                                         & 10.7\%                                                                             & unboosted                                                           \\
Local adaptivity              &  \cite{songetal}  & 6ring                   & 6 ring                                                                          & \{5, 3\}                                                           & 90                                                                                         & 11.5\%                                                                             & unboosted                                                           \\
Local adaptivity              &  \cite{songetal}  & 6ring                   & 6 ring                                                                          & \{6, 3\}                                                           & 108                                                                                        & 11.9\%                                                                             & unboosted                                                           \\
Local adaptivity              &  \cite{songetal}  & 6ring                   & 6 ring                                                                          & \{5, 4\}                                                           & 120                                                                                        & 12.5\%                                                                             & unboosted                                                           \\
Local adaptivity              &  \cite{songetal}  & 6ring                   & 6 ring                                                                          & \{6, 4\}                                                           & 144                                                                                        & 13.3\%                                                                             & unboosted                                                           \\
Local adaptivity              &  \cite{songetal}  & 6ring                   & 6 ring                                                                          & \{7, 4\}                                                           & 168                                                                                        & 14.0\%                                                                             & unboosted                                                           \\
Local adaptivity              &  \cite{songetal}  & 6ring                   & 6 ring                                                                          & \{10, 4\}                                                          & 240                                                                                        & 14.0\%                                                                             & unboosted                                                           \\
Local adaptivity              &  \cite{songetal}  & 6ring                   & 6 ring                                                                          & \{9, 6\}                                                           & 324                                                                                        & 14.0\%                                                                             & unboosted                                                           \\
Local adaptivity              &  \cite{songetal}  & 6ring                   & 6 ring                                                                          & \{17, 4\}                                                          & 408                                                                                        & 12.4\%                                                                             & unboosted                                                           \\
Local adaptivity              &  \cite{songetal}  & 6ring                   & 6 ring                                                                          & \{12, 7\}                                                          & 504                                                                                        & 13.9\%                                                                             & unboosted                                                           \\
Statis bias arrangement       &         & 6ring                   & 6 ring                                                                          & \{2, 3\}                                                           & 36                                                                                         & 5.1\%                                                                              & unboosted                                                           \\
Statis bias arrangement       &            & 6ring                   & 6 ring                                                                          & \{3, 4\}                                                           & 72                                                                                         & 7.7\%                                                                              & unboosted                                                           \\
Statis bias arrangement       &             & 6ring                   & 6 ring                                                                          & \{4, 7\}                                                           & 168                                                                                        & 11.3\%                                                                             & unboosted                                                           \\
Statis bias arrangement       &            & 6ring                   & 6 ring                                                                          & \{5, 20\}                                                          & 600                                                                                        & 16.7\%                                                                             & unboosted                                                           \\
Statis bias arrangement       &            & 6ring                   & 6 ring                                                                          & \{7, 100\}                                                         & 4200                                                                                       & 20.8\%                                                                             & unboosted                                                           \\
Statis bias arrangement       &           & 6ring                   & 6 ring                                                                          & \{10, 1000\}                                                       & 60000                                                                                      & 23.6\%                                                                             & unboosted                                                           \\
Statis bias arrangement       &            & 6ring                   & 6 ring                                                                          & \{13, 10000\}                                                      & 780000                                                                                     & 24.9\%                                                                             & unboosted                                                           \\
Statis bias arrangement       &            & 6ring                   & 6 ring                                                                          & \{16, 100000\}                                                     & 9600000                                                                                    & 25.6\%                                                                             & unboosted                                                           \\
Randomized failure            & \cite{FBQC}     & 6ring                   & 6 ring                                                                          & \{2, 2\}                                                           & 24                                                                                         & 2.7\%                                                                              & boosted                                                             \\
Randomized failure            & \cite{FBQC}      & 6ring                   & 6 ring                                                                          & \{2, 3\}                                                           & 36                                                                                         & 3.5\%                                                                              & boosted                                                             \\
Randomized failure            & \cite{FBQC}      & 6ring                   & 6 ring                                                                          & \{3, 4\}                                                           & 72                                                                                         & 4.8\%                                                                              & boosted                                                             \\
Randomized failure            & \cite{FBQC}      & 6ring                   & 6 ring                                                                          & \{4, 7\}                                                           & 168                                                                                        & 5.9\%                                                                              & boosted                                                             \\
Randomized failure            & \cite{FBQC}       & 6ring                   & 6 ring                                                                          & \{5, 20\}                                                          & 600                                                                                        & 8.1\%                                                                              & unboosted                                                           \\
Randomized failure            & \cite{FBQC}       & 6ring                   & 6 ring                                                                          & \{7, 100\}                                                         & 4200                                                                                       & 11.0\%                                                                             & unboosted                                                           \\
Randomized failure            & \cite{FBQC}       & 6ring                   & 6 ring                                                                          & \{10, 1000\}                                                       & 60000                                                                                      & 13.1\%                                                                             & unboosted                                                           \\
Randomized failure            & \cite{FBQC}      & 6ring                   & 6 ring                                                                          & \{13, 10000\}                                                      & 780000                                                                                     & 13.9\%                                                                             & unboosted                                                           \\
Randomized failure            & \cite{FBQC}       & 6ring                   & 6 ring                                                                          & \{16, 100000\}                                                     & 9600000                                                                                    & 14.3\%                                                                             & unboosted                                                           \\ \bottomrule
\end{tabular}
\caption{Loss per photon threshold data. Where local encodings are described as \{n,m\} they are Shor codes (also known as quantum parity codes) with n- X repetitions and m- Z repetitions. Local encodings in~\cite{bell2023optimizing} are n-qubit optimized graph codes. From \cite{pankovich} we include the highest performing reuslts of the ``active+cyclic" architecture. From \cite{bell2023optimizing} we include the best performing adaptive measurement results for each code size. }
\label{table:data}
\end{table*}

\end{document}